\documentclass[aps,twocolumn,groupedaddress,superscriptaddress,floatfix,showpacs,amsmath,amssymb,prb]{revtex4}
\usepackage{amsmath}
\usepackage{amssymb}
\usepackage{graphicx}
\usepackage{textcomp}
\usepackage{bm}	

\usepackage[usenames,dvipsnames]{color}	

\begin{document}
    \title{Steady-state entanglement between distant quantum dots in photonic crystal dimers}
    
\author{J. P. Vasco}\email{jpvasco@gmail.com}
\affiliation{Departamento de F\'{\i}sica, Universidade Federal de Minas Gerais, Belo Horizonte MG, Brazil}
\affiliation{DISSE - INCT de Nanodispositivos Semicondutores, Brazil} 
\author{D. Gerace}\email{dario.gerace@unipv.it}    
\affiliation{Dipartimento di Fisica, Universit\`a di Pavia, via Bassi 6, I-27100 Pavia, Italy}
\author{P. S. S. Guimar\~{a}es}\email{pssg@fisica.ufmg.br}
\affiliation{Departamento de F\'{\i}sica, Universidade Federal de Minas Gerais, Belo Horizonte, Minas Gerais, Brazil}
\affiliation{DISSE - INCT de Nanodispositivos Semicondutores, Brazil} 
\author{M. F. Santos}\email{mfsantos@if.ufrj.br}
\affiliation{Departamento de F\'{\i}sica, Universidade Federal de Minas Gerais, Belo Horizonte MG, Brazil}
\affiliation{Instituto de F\'{\i}sica, Universidade Federal do Rio de Janeiro, Rio de Janeiro, Brazil}

\begin{abstract} 
We show that two spatially separated semiconductor quantum dots under resonant and continuous-wave excitation can be strongly entangled in the steady-state, thanks to their radiative coupling by mutual interaction through the normal modes of a photonic crystal dimer. We employ a quantum master equation formalism to quantify the steady-state entanglement by calculating the system {\it negativity}.
Calculations are specified to consider realistic semiconductor nanostructure parameters for the photonic crystal dimer-quantum dots coupled system, determined by a guided mode expansion solution of Maxwell equations. Negativity values of the order of 0.1 ($20\%$ of the maximum value) are shown for interdot distances that are larger than the resonant wavelength of the system. It is shown that the amount of entanglement is almost independent of the interdot distance, as long as the normal mode splitting of the photonic dimer is larger than their linewidths, which becomes the only requirement to achieve a local and individual qubit addressing.
Considering inhomogeneously broadened quantum dots, we find that the steady-state entanglement is preserved as long as the detuning between the two quantum dot resonances is small when compared to their decay rates. The steady-state entanglement is shown to be robust against the effects of pure dephasing of the quantum dot transitions. We finally study the entanglement dynamics for a configuration in which one of the two quantum dots is initially excited and find that the transient negativity can be enhanced by more than a factor of two with respect to the steady-state value. These results are promising for practical applications of entangled states at short time scales.
\end{abstract}

\maketitle

\section{Introduction}

The possibility of exploiting modern semiconductor devices for quantum information processing has attracted considerable attention in the past decade \cite{imamoglu}. In particular, semiconductor quantum dots (QDs) are among the best candidates for quantum bit (qubit) operations, owing to their unique atom-like radiative properties combined with the sophisticated integration techniques achieved nowadays on semiconductor platforms \cite{warburton}. 
Entangling distinct quantum emitters is a key requirement for such applications, both for quantum gate engineering and quantum information transferring \cite{veldhorst,shim}. With this aim, the coherent interaction between two QDs has been widely studied in the short-distance regime \cite{nazir,borges1,borges2}, and recent experimental evidence has shown their actual relevance for quantum computing \cite{bayer,gao}. However, individual qubit manipulation, which is crucial for practical applications, remains challenging due to the small spatial separations typically needed for achieving strongly entangled QD states. On a parallel ground, solid-state artificial atoms inevitably suffer from short coherence times owing to their coupling to dissipative environment \cite{lodahlrev}, making it difficult to envision long-lived entangled states with such kind of qubits for practical applications. To date, a conclusive demonstration of long-lived entangled states in spatially separated and distant QDs has not been shown.  

To this end, a number of studies have been recently addressing the mutual QDs coupling that is indirectly mediated by purely photonic degrees of freedom \cite{parascandolo,tarel,hughes1}. The main goal would be to achieve a long-range coherent energy transfer between the QDs by overcoming their short-distance interactions, such as tunneling and F\"{o}rster coupling, thus enabling individual qubit manipulation. Among the different nanophotonic systems, photonic crystals represent one of the most promising platforms to achieve such long-range interactions between spatially separated QDs \cite{hughes2,minkov2013a,minkov2013b}. Thanks to the enormous progress in fabrication technologies, QDs coupled to photonic crystal cavities have allowed pioneering demonstrations of cavity quantum electrodynamics phenomena, such as Purcell enhancement, lasing, and strong light-matter coupling at the single quantum level \cite{kress2005,englund2005,ellis,yoshie,kevin07nat,reinhard2012nphot,somaschi2016}. It has been proposed that two QDs can be entangled through coupling within the same cavity \cite{imamoglu2007,gallardo,villasboas,Elena,albert}, although the interdot distance remains smaller than the operational wavelength. 
Based on these grounds, it has been recently proposed that sizable radiative coupling can be achieved between QDs embedded in photonic crystal dimers (PCD) \cite{jpvasco1,jpvasco2}, i.e., coupled photonic crystal cavities, where the QDs distance can be considerably larger than their resonant wavelength.
Such a system could be useful to finally obtain a long-distance entanglement, an alternative to recent proposals based on plasmonic \cite{gonzalez2011prl,prbtudela2011,hughes3} or nanowire photonic crystal \cite{hughes4} approaches. 
Moreover, the PCD approach might be promising to realize a recently proposed scheme allowing us to obtain steady-state entanglement via quantum bath engineering \cite{AronPRA2014}, thus overcoming the known issues of QDs in terms of short coherence times.

\begin{figure}[t]
  \begin{center}
    \includegraphics[width=0.45\textwidth]{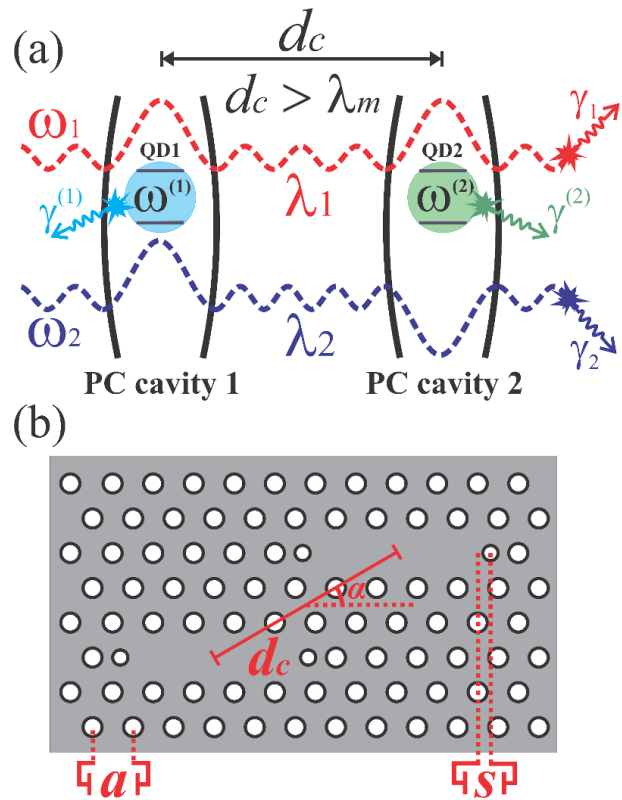}
  \end{center}
  \caption{
(Color online) (a) Schematic representation of the system studied in this work. Two strongly coupled cavities with one quantum dot coupled to the electric field antinode in each of them. The normal modes, arising from the hybridization of the fundamental cavity mode in each cavity, radiatively couple the quantum dots. The $\omega$'s represent the frequencies of the system while the $\gamma$'s represent the loss rates. The intercavity distance $d_c$, which also defines the dot-dot separation, can be larger than the characteristic wavelength of the system, $\lambda_m$. (b) PCD considered in this work; two strongly coupled L3 photonic crystal cavities in a hexagonal lattice of holes with lattice parameter $a$. The end lateral holes are displaced by $s$ outward, and their radii are decreased to $80\%$ of the nominal value. The angle between the line connecting the centers of the cavities and the horizontal axis is $\alpha$.}
  \label{fig:system}
\end{figure}

In this work we propose to exploit the delocalized nature of the normal modes of the PCD and their strong dipole coupling to QD excitons to simultaneously achieve a sizable entanglement between two distant solid-state qubits and in the steady-state. Figure~\ref{fig:system}(a) shows a schematic illustration of our proposed system: Two strongly coupled photonic crystal cavities (the PCD) mediate the coupling between two QDs positioned at their respective electric-field antinodes through the electromagnetic normal modes of the coupled cavity system. Unlike a recent proposal \cite{AronPRA2014}, we hereby assume to resonantly and continuously drive the QD excitons directly, e.g. via coherent electrical $\pi$-pulses \cite{zrenner2002Nat}, and show that this is sufficient to achieve a sizable steady-state entanglement between the two spatially separated qubits even when considering photonic and excitonic losses.  
The present paper complements our recent study \cite{jpvasco1,jpvasco2} on the coherent energy transfer between two distant QDs in PCDs, where we used a semiclassical approach based on the photonic Green's function. Here we employ a fully quantum mechanical theory, where we focus on the role of the normal modes as channels for quantum entanglement between the QDs. The present study will be useful for quantum information applications on a photonic crystal chip.

The work is organized as follows. In Sec.~\ref{Theory}, we summarize the semiclassical approach used to estimate the photonic crystal parameters for realistic and state-of-art coupled QD-PCD configurations and materials, and the quantum mechanical model of two QDs coupled via the normal modes of the PCD using the master equation formalism. The steady-state entanglement between the QDs as a function of the interdot distance is studied in Sec.~\ref{SSEnt} for all possible geometric configurations of the PCD considered in this work. In Sec.~\ref{DyEnt} we study the transient dynamics of the system and we propose a simple approach for generation of strongly entangled QDs in practical applications. Finally, the main conclusions of the work are presented in Sec.~\ref{Concl}.

\section{Theory and methods }\label{Theory}

In order to describe the system that is schematically sketched in Fig.~\ref{fig:system}(a), we apply a two-step theoretical approach: First, the relevant system parameters are estimated in a practical realization by numerically solving Maxwell equations in a realistic photonic crystal nanostructure, and then these theoretically derived parameters (such as QD-cavity coupling and dissipation rates) are used as inputs for a quantum master equation formalism, which allows us to quantify the degree of entanglement between the two qubits. The two theories are briefly outlined below.

\textit{Guided mode expansion}. Among the practical realizations of the model system outlined in Fig.~\ref{fig:system}(a) on an integrated nanophotonics platform, we are specifically interested in describing the PCD formed by two coupled (nominally identical) L3 photonic crystal slab cavities in a hexagonal lattice of circular holes as shown in Fig.~\ref{fig:system}(b). The L3 cavity consists in three missing holes along the $\Gamma K$ lattice direction \cite{nodaL3}, and we adopt the optimized design in which the end lateral holes are displaced by 0.15$a$ outward, $a$ being the lattice constant of the underlying photonic crystal lattice, and their radii are decreased to $80\%$ of the nominal value \cite{gerace04pnfa}. The geometry of the photonic crystal lattice allows four possible symmetrical alignments between the L3 cavities; i.e., the line connecting the centers of the cavities can determine angles $\alpha$ of $0^{\circ}$, $30^{\circ}$, $60^{\circ}$, and $90^{\circ}$ with respect to the largest cavity axis \cite{chalcraftOpex}. The two normal mode frequencies and their respective loss rates (quality factors), arising from hybridization of the L3 fundamental cavity modes in the neighboring cavities, are calculated by using the guided mode expansion method (GME), in which the electromagnetic fields of the photonic crystal slab are expanded in the guided mode basis of the equivalent homogeneous planar waveguide \cite{andreani06prb}. We assume that the point-like QDs are positioned at the electric field antinodes, i.e., in the centers of the two L3 cavities, where E$_y$ is the only non-vanishing electric field component \cite{chalcraftapl}. At this optimal condition the QD-field coupling strengths can be written as \cite{minkov2013a,minkov2013b}:
\begin{equation}
 g_m^{(n)}=\left(\frac{2\pi\omega_0 d^2}{\hbar}\right)^{1/2} \mbox{E}_{y,m}(\mathbf{r}_n),
\end{equation}
where $\omega_0$ and $d^2$ are an average exciton transition frequency and the squared QD dipole moment, which are in the range of $\sim 1.3$~eV and $\sim 0.51$~eV$\cdot$nm$^3$, respectively, for typical self-assembled InGaAs QDs \cite{kevin07nat,reinhard2012nphot}; $\mathbf{r}_n$ is the position of the QD $n$, and the electric field of the normal mode $m$ is subject to the normalization condition $\int \epsilon(\mathbf{r})\mathbf{E}_m^{\ast}(\mathbf{r})\cdot\mathbf{E}_m(\mathbf{r})d\mathbf{r}=1$. The normal mode frequencies, $\omega_1$ and $\omega_2$, and the coupling strengths $g_m^{(n)}$ are implicit functions of the distance between the cavities, $d_c$, defined as the center-to-center distance which thus coincides with the distance between the QDs. We chose system parameters relevant to III-V GaAs-based structures, i.e., lattice constant $a=260$~nm, hole radius $65$~nm, slab thickness $120$~nm and real part of the refractive index $\sqrt{\epsilon_\infty}=3.41$. For computing the loss rates of the normal modes, $\gamma_m$, we adopt the photonic Fermi's golden rule into the GME approximation, in which the transition rate from a guided mode to a leaky mode is calculated using time-dependent perturbation theory \cite{sakodaprb,andreani06prb}. Since the coupling between the guided modes and the radiation modes depends on the near field distribution throughout the dielectric structure \cite{jelenafarfield,bonatofarfield}, the loss rates change when the distance between the cavities varies, and the dependence of losses on $d_c$ is usually strongest for small intercavity distances. Therefore, $\gamma_m$ is also an implicit function of $d_c$.

The numerical GME calculations for computing the relevant system parameters, namely $\omega_m$, $\gamma_m$, and $\mbox{E}_{y,m}$, are carried out using a hexagonal supercell of superlattice parameter $24a$ for the $30^{\circ}$ and $60^{\circ}$ cases, with 11025 plane waves tested for convergence. Rectangular supercells of dimensions $27a\times8\sqrt{3}a$ with 11915 plane waves, tested for convergence, and $18a\times25\sqrt{3}a$ with 24829 plane waves, tested for convergence, are used for the $0^{\circ}$ and $90^{\circ}$ cases, respectively. Only one guided mode is used in the guided mode expansion, since the corrections of high order guided modes are negligible for the slab thickness and the refractive index considered in this work.

\textit{Master equation formalism}. 
The system of two QDs coupled to the normal modes of a PCD can be described by a second-quantized Hamiltonian, written in the normal mode basis, where the rotating wave approximation is employed and the QDs are assumed as point-like two level systems:
\begin{equation}\label{Ham}
\begin{aligned}
\hat{H}= & \sum_{m=1}^{2}\left(\hbar\omega_m\hat{a}^{\dag}_m\hat{a}_m+\hbar\omega^{(m)}\hat{\sigma}^{+}_m\hat{\sigma}^{-}_m\right)\\
 & + \sum_{m,n=1}^{2}\left(\hbar g_m^{\ast(n)}\hat{a}^{\dag}_{m}\hat{\sigma}^{-}_n+\hbar g_m^{(n)}\hat{a}_m\hat{\sigma}^{+}_n\right)\\
 & + \sum_n^{2}\left(\hbar\Omega_n e^{-i\omega_pt}\hat{\sigma}^{+}_n+\hbar\Omega_n^{\ast} e^{i\omega_pt}\hat{\sigma}^{-}_n\right).\\
\end{aligned}
\end{equation}
Here, $\omega_m$ and $\omega^{(m)}$ correspond to the frequency of the normal mode $m$ and the excitonic transition frequency of the QD $m$, respectively; $\hat{a}^{\dag}_{m}$ ($\hat{a}_{m}$) is the creation (destruction) operator of photons in the normal mode $m$; $\hat{\sigma}^{+}_m$ ($\hat{\sigma}^{-}_m$) is the creation (destruction) operator of one electron-hole pair in the QD $m$; $g_m^{(n)}$ are the coupling strengths between normal mode $m$ and QD $n$; and $\Omega_n$ is the pumping rate at which electron-hole pairs in the QD $n$ are coherently created by a continuous-wave pump laser, or electric gating potential, with frequency $\omega_p$. With the aim of eliminating the explicit temporal dependence in the Hamiltonian of Eq.~(\ref{Ham}), the system dynamics can be described in a rotating reference frame with frequency $\omega_p$ by applying the operator $\hat{R}(t)=\mbox{exp}\left[i\omega_pt\left(\hat{a}^{\dag}_1\hat{a}_1+\hat{a}^{\dag}_2\hat{a}_2+\hat{\sigma}^{+}_1\hat{\sigma}^{-}_1+\hat{\sigma}^{+}_2\hat{\sigma}^{-}_2\right)\right]$, determining an effective Hamiltonian $\hat{H}_{\mbox{eff}}=\hat{R}\hat{H}\hat{R}^{\dag}-i\hbar\hat{R}\left(d\hat{R}^{\dag}/dt\right)$, i.e.,

\begin{equation}\label{Hamrot}
\begin{aligned}
\hat{H}_{\mbox{eff}}= & \sum_{m=1}^{2}\left(\hbar\bar{\omega}_m\hat{a}^{\dag}_m\hat{a}_m+\hbar\bar{\omega}^{(m)}\hat{\sigma}^{+}_m\hat{\sigma}^{-}_m\right)\\
 & + \sum_{m,n=1}^{2}\left(\hbar g_m^{\ast(n)}\hat{a}^{\dag}_{m}\hat{\sigma}^{-}_n+\hbar g_m^{(n)}\hat{a}_m\hat{\sigma}^{+}_n\right)\\
 & + \sum_n^{2}\left(\hbar\Omega_n\hat{\sigma}^{+}_n+\hbar\Omega_n^{\ast}\hat{\sigma}^{-}_n\right),\\
\end{aligned}
\end{equation}

\noindent where $\bar{\omega}_m=\omega_m-\omega_p$ and $\bar{\omega}^{(m)}=\omega^{(m)}-\omega_p$. We adopt the master equation formalism to quantitatively account for the losses of the system, which is written in Markov approximation for the rotated density matrix, i.e., $\tilde{\rho}=\hat{R}\rho\hat{R}^{\dag}$, as:

\begin{equation}\label{master}
 \frac{d\tilde{\rho}}{dt}=\frac{i}{\hbar}\left[\tilde{\rho},\hat{H}_{\mbox{eff}}\right]+\sum_{m=1}^{2}\left[{\cal L}(\gamma_m)+{\cal L}(\gamma^{(m)})\right],
\end{equation}

\noindent where ${\cal L}(\gamma_m)=\gamma_m\left[\hat{a}_m\tilde{\rho}\hat{a}^{\dag}_m-\hat{a}^{\dag}_m\hat{a}_m\tilde{\rho}/2-\tilde{\rho}\hat{a}^{\dag}_m \hat{a}_m/2\right]$ and ${\cal L}(\gamma^{(m)})=\gamma^{(m)}\left[\hat{\sigma}^{-}_m\tilde{\rho}\hat{\sigma}^{+}_m-\hat{\sigma}^{+}_m \hat{\sigma}^{-}_m \tilde{\rho}/2-\tilde{\rho}\hat{\sigma}^{+}_m\hat{\sigma}^{-}_m/2\right]$ are the Lindblad operators corresponding to the losses (both intrinsic and extrinsic, respectively) of the photonic normal mode $m$ at a rate $\gamma_m$, as well as the losses by spontaneous emission in the QD $m$ at an exciton decay rate $\gamma^{(m)}$. The former are explicitly calculated for the PCD nanostructure by the GME approach described above. Moreover, since we are interested in low excitation powers, low temperature and resonant excitation regimes, the Lindblad dissipation terms associated with incoherent pumping can be safely neglected in our master equation model.
Pure dephasing of the QDs transitions can be taken into account by an additional Lindblad term
${\cal L}_d(\gamma_d^{(m)})=\gamma_d^{(m)}\left[\hat{\sigma}^{+}_m\hat{\sigma}^{-}_m\tilde{\rho}\hat{\sigma}^{+}_m\hat{\sigma}^{-}_m-(\hat{\sigma}^{+}_m \hat{\sigma}^{-}_m)^2 \tilde{\rho}/2-\tilde{\rho}(\hat{\sigma}^{+}_m\hat{\sigma}^{-}_m)^2/2\right]$
where $\gamma_d^{(m)}$ represents a pure dephasing rate.
The master equation is numerically implemented by expressing the operators on an occupation number Fock basis, truncated to the most suitable photon number previously checked for convergence.
 
We are ultimately interested in quantifying the entanglement between the two QDs as mutually coupled qubits, for which we employ the Peres-Horodecki negativity criterion \cite{peres,horodecki,karolneg}. The latter accounts for the non-separability condition of the reduced density matrix in the composite Hilbert space of dimension $2\otimes2$, effectively describing the two qubits quantum mechanical behavior. The negativity is an entanglement monotone for a two-qubit system which, for the hereby used normalization, ranges from zero for a separable state up to the maximum value $0.5$ for the maximally entangled Bell states (see the appendix~\ref{appx} for details).
In our case, the reduced density matrix of the QDs is numerically calculated by tracing over the photonic normal modes $\rho_{QD1QD2}=\mbox{Tr}\left[\rho\right]_m$, and the negativity, quantifying the degree of entanglement between the QDs, is defined as the absolute value of the sum of the negative eigenvalues of $\rho_{QD1QD2}^{T1}$, where $T1$ represents the partial transpose of $\rho_{QD1QD2}$ with respect to the system 1, i.e., QD 1.

\section{Steady-state entanglement}\label{SSEnt}

\begin{figure}[t]
  \begin{center}
    \includegraphics[width=0.45\textwidth]{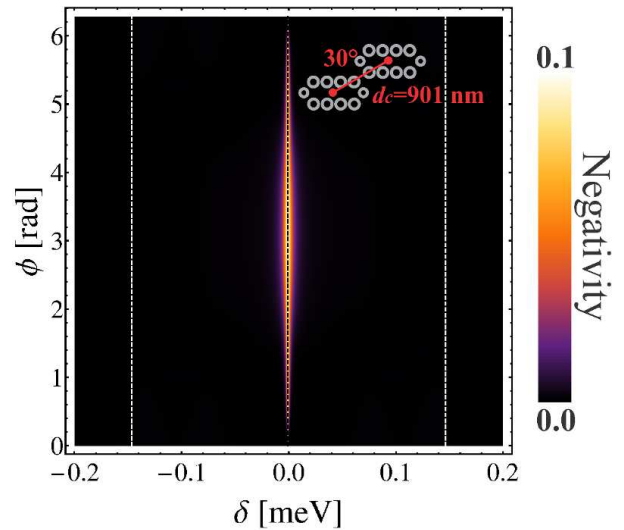}
  \end{center}
  \caption{
(Color online) Steady-state negativity for the $30^{\circ}$ dimer at $d_c=2\sqrt{3}a=901$~nm as a function of the phase difference between the pumpings $\phi=\phi_1-\phi_2$, with $\phi_2=0$, and the frequency shift $\delta$, where $\omega_p=\omega_1+\delta$. The largest negativity is 0.103 or $\sim 20\%$ of the maximum value. The two QDs are in resonance with the lower frequency normal mode and we have considered $\gamma^{(m)}=0$ and $\hbar\Omega_0=1$~$\mu$eV. The vertical black and white dashed lines correspond to the dark state and polariton branches of the system, respectively.}
  \label{fig:30-dplot}
\end{figure}

\begin{figure*}[hpt!]
  \begin{center}
    \includegraphics[width=0.8\textwidth]{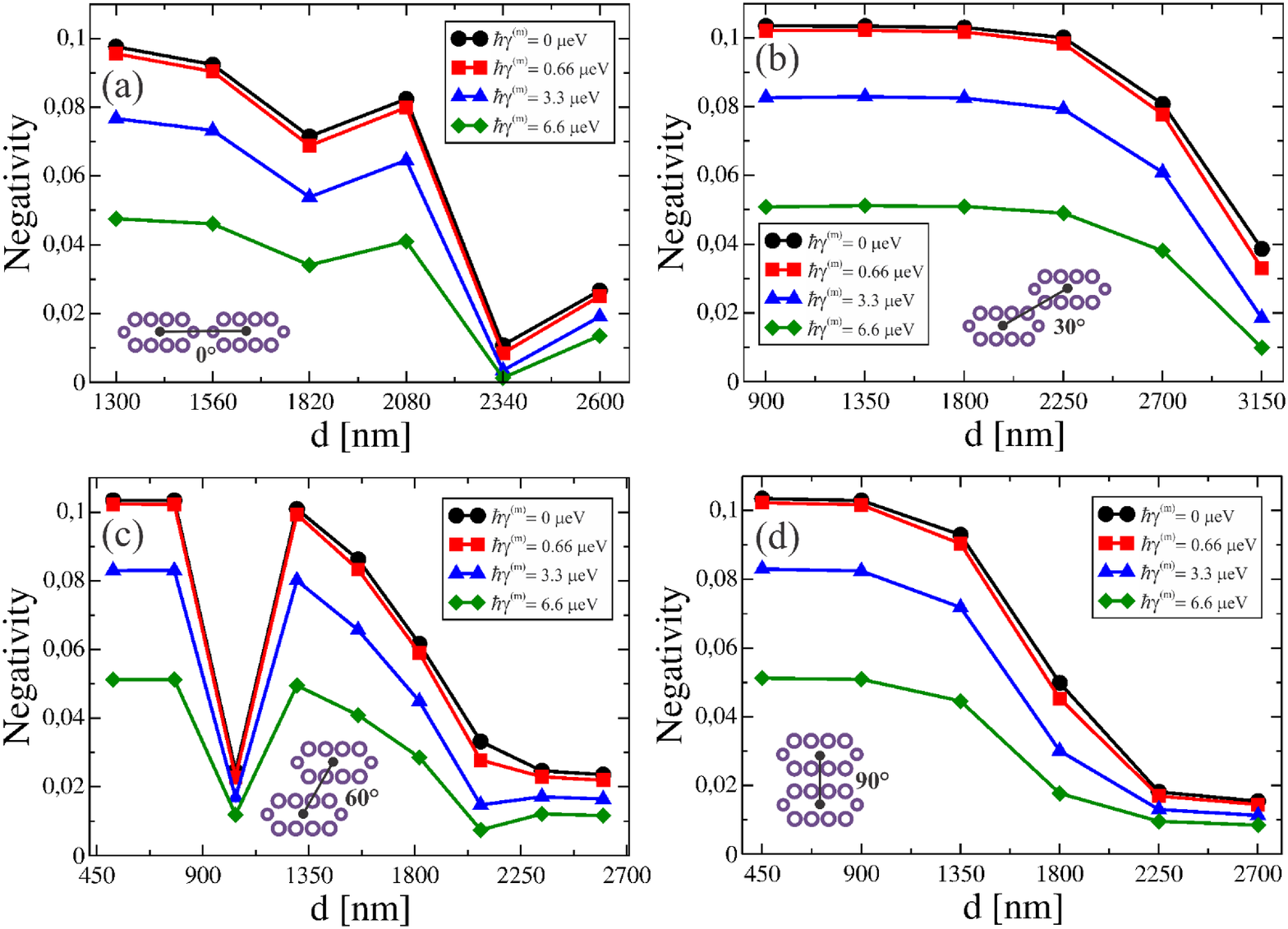} 
  \end{center}
  \caption{
(Color online) Steady-state negativity for the dimers with the connecting lines at (a) $0^{\circ}$, (b) $30^{\circ}$, (c) $60^{\circ}$, and (d) $90^{\circ}$, as a function of the distance between the QDs, for different values  $\hbar\gamma^{(m)}=0$~$\mu$eV, black circles; $\hbar\gamma^{(m)}=0.66$~$\mu$eV, red squares; $\hbar\gamma^{(m)}=3.3$~$\mu$eV, blue triangles; and $\hbar\gamma^{(m)}=6.6$~$\mu$eV, green diamonds. The two QDs are in resonance with the lower frequency normal mode and $\hbar\Omega_0=1$~$\mu$eV. The lines connecting the individual points only serve as a guide to the eye.}
\label{fig:neg-vs-d}
\end{figure*}

In this section, we are interested in characterizing the photonic normal modes as channels for quantum entanglement in the steady-state. For determining the state with maximum entanglement we write the pumping rates of the QDs in the form $\Omega_n=\Omega_0e^{i\phi_n}$, where $\phi=\phi_1-\phi_2$ is their phase difference, and we write the pumping frequency as $\omega_p=\omega_1+\delta$. Considering the two QDs resonant with the lower frequency normal mode, $\omega_1$, and using the calculated GME parameters in the quantum model, we compute the negativity by solving the master equation for the steady-state density matrix as a function of $\phi$ and $\delta$, with $\phi_2=0$. Figure~\ref{fig:30-dplot} shows the results for the $30^{\circ}$ PCD at distance $d_c=2\sqrt{3}a=901$~nm, where the vertical black and white dashed lines correspond to the dark state and polariton branches of the system, respectively. The largest entanglement, corresponding to a negativity of 0.103 (or $\sim 20\%$ of the maximum value, see the appendix), is seen for a pumping frequency that is resonant with the dark state, and a $\pi$ phase difference. Since the excitonic dark state does not couple effectively to the photonic mode, due to their opposite symmetry, the former remains ``protected'' from the dissipative effects of the latter, which allows for the non-zero steady-state negativity. The phase difference between the pumpings is determined by the bonding (symmetric) or antibonding (antisymmetric) character of the normal mode; for bonding modes the optimal phase difference will be $(2n+1)\pi$ (antisymmetric excitonic dark state) while for antibonding modes it will be $2n\pi$ (symmetric excitonic dark state), with $n$ integer. In the calculations of Fig.~\ref{fig:30-dplot} we have considered $\gamma^{(m)}=0$ and $\hbar\Omega_0=1$~$\mu$eV; since $\hbar\gamma_m$ is between 10 and 60~$\mu$eV and $\hbar g_m^{(n)} \sim 110$~$\mu$eV for all dimers, we are in the weak pumping regime and the basis used for solving the master equation, $|\alpha_1\alpha_2m_1m_2\rangle$ is truncated at $m_i=1$ (we have checked that it is sufficient for convergence), where $\alpha_i=0$ or 1 is the excitation number in the QD $i$, and $m_i$ is the number of photons in the mode $i$. We have obtained equivalent results for all of the PCD configurations considered, and at all inter-cavity distances allowed by the corresponding supercell. We have also verified in our calculations that if we pump coherently only the photonic mode, the steady-state entanglement of the excitonic dark state is destroyed. This is due to the opposite symmetry between the excitonic dark state and the resonant photonic mode, which means that the optimal condition to entangle the QDs is not fulfilled when only the photonic mode is coherently pumped.

\begin{figure}[pt!]
  \begin{center}
    \includegraphics[width=0.45\textwidth]{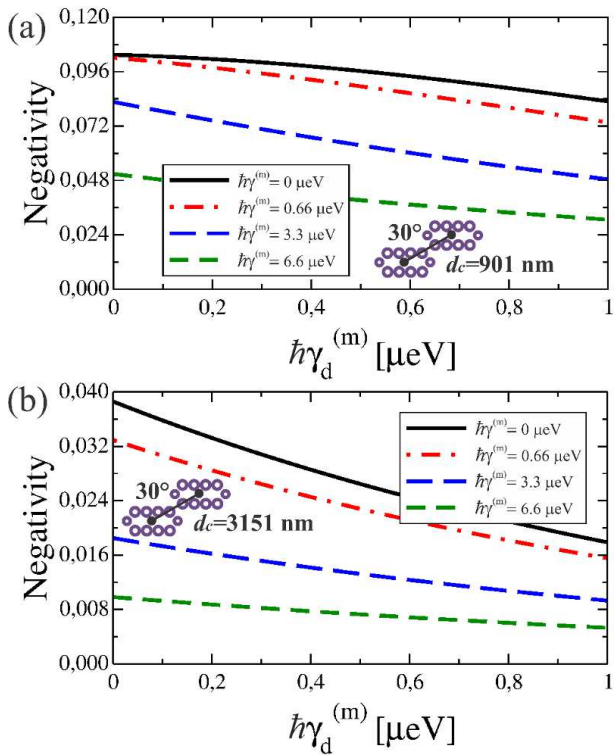} 
  \end{center}
  \caption{
(Color online) Steady-state negativity for the $30^{\circ}$ dimer at $d_c=2\sqrt{3}a=901$~nm, panel (a), and $d_c=7\sqrt{3}a=3151$~nm, panel (b), as a function of the pure dephasing rate $\gamma_d^{(m)}=\gamma_d^{(1)}=\gamma_d^{(2)}$, for different QD loss rates. The two QDs are in resonance with the lower frequency normal mode, i.e., $\omega^{(1)}=\omega^{(2)}=\omega_1$, and the pumping rate is $\hbar\Omega_0=1$~$\mu$eV.}
\label{fig:neg-vs-deph}
\end{figure}

Taking into account that the largest steady-state negativity corresponds to the dark state, for a pump's phase difference determined by the photonic mode in resonance with the QDs, we now investigate how the entanglement depends on the QDs separation, $d_c$. Figure~\ref{fig:neg-vs-d} shows the negativity calculated as a function of the interdot distance, for the $0^{\circ}$, $30^{\circ}$, $60^{\circ}$, and $90^{\circ}$ dimers, considering $\hbar\gamma^{(m)}=0$~$\mu$eV, black circles; $\hbar\gamma^{(m)}=0.66$~$\mu$eV, red squares; $\hbar\gamma^{(m)}=3.3$~$\mu$eV, blue triangles; and $\hbar\gamma^{(m)}=6.6$~$\mu$eV, green diamonds. The QDs are in resonance with the lower frequency normal mode and $\hbar\Omega_0=1$~$\mu$eV as in Fig.~\ref{fig:30-dplot}. We find that the negativity decreases as a function of the interdot distance in the large $d_c$ region for all dimers, which suggests a direct proportionality between the QDs entanglement and the PCD normal mode splitting. The latter is known to be a decreasing monotonic function for large intercavity distances \cite{chalcraftapl,jpvasco1}. On the other hand, in PCD the normal mode splitting is not monotonic for intermediate values of $d_c$; in fact, the splitting can increase for increasing intercavity distance at specific PCD configurations \cite{chalcraftOpex}. Such a phenomenon is clearly reflected in the negativity, i.e., the entanglement increases for increasing $d_c$, into the $d_c$ intervals [1820,2080]~nm  and [2340,2600]~nm for Fig.~\ref{fig:neg-vs-d}(a), and [1040,1300]~nm for Fig.~\ref{fig:neg-vs-d}(c); in these cases, the normal mode splitting changes from a very small value to a large value, with respect to the linewidths of the photonic modes. At the other intermediate values of interdot distances, the negativity is roughly of the order of $\sim 0.1$, i.e., $\sim 20\%$ of the maximum value. Hence, the results of Fig.~\ref{fig:neg-vs-d} show that the negativity remains of the order of $\sim 0.1$ as long as the normal mode splitting is spectrally well-defined (i.e., larger than the photonic linewidths), which is actually the regime where the effective dipole-dipole interaction is proportional to the quality factor of the resonant normal mode, as extensively investigated in a previous work \cite{jpvasco1}. The $30^{\circ}$ dimer, in Fig.~\ref{fig:neg-vs-d}(b), clearly evidences such a behavior; the negativity is a very flat function, around $0.1$, up to $d_c=2252$~nm, where the mode splitting is much larger than the normal mode linewidths. For larger values of $d_c$, the splitting becomes of the order of $\gamma_m$ and the negativity decreases. Owing to the lower penetration into the photonic crystal barriers for the L3 cavity modes along the cavity axis, the $90^{\circ}$ PCD is characterized by a rapidly decreasing normal mode splitting on increasing $d_c$, as it is evident in Fig.~\ref{fig:neg-vs-d}(d). As a consequence, significant values of negativity are not supported at interdot distances that are larger than the characteristic wavelength of the system. Furthermore, it is very interesting that the entanglement is not strongly affected by the $\gamma_m$ rates as long as the normal mode splitting is well-defined; along the flat region (negativity almost independent on interdot distance) of the $30^{\circ}$ PCD, $\hbar\gamma_1$ and $\hbar\gamma_2$ change from $67$ and $37$~$\mu$eV, to $17$ and $16$~$\mu$eV, respectively, when $d_c$ correspondingly changes from 901 to 2252~nm. From a previous study, it is known that the resonant energy transfer between radiatively-coupled QDs depends on the quality factor of the normal mode in resonance with the dots, where the $0^{\circ}$ configuration is the most convenient in terms of energy transfer, due to its very high normal mode quality factors \cite{jpvasco2}. In the present work we essentially show that when the relevant figure of merit is the long-range entanglement, the $30^{\circ}$ dimer is the best choice due to its well-defined normal mode splitting even for distances larger than the characteristic wavelength of 
the system. 

The results of Fig.~\ref{fig:neg-vs-d} also evidence that the entanglement of the dark state is only marginally affected when losses of typical self-organized InGaAs QDs are taken into account; state-of-art InGaAs QD excitonic lifetimes are between $0.2$ and $1$~ns, as experimentally reported in the literature \cite{majumdar2012,reinhard2012nphot,somaschi2016}. As a further loss channel, semiconductor QDs are known to be subject to pure dephasing \cite{hughes5,villasboas}. To complete the study on the dependence of entanglement on the main system losses, in Fig.~\ref{fig:neg-vs-deph} we investigate the dependence of the steady-state negativity on their pure dephasing rates. Results are reported for the $30^{\circ}$ dimer at two different interdot distances in Figs.~\ref{fig:neg-vs-deph}(a) and \ref{fig:neg-vs-deph}(b), and considering the same values of $\gamma^{(m)}$ rates as in Fig.~\ref{fig:neg-vs-d}. The steady-state entanglement is not strongly affected by viable experimental pure dephasing rates \cite{hughes4}. In Fig.~\ref{fig:neg-vs-deph}(a), where the normal mode splitting is much larger than the photonic linewidth, the negativity is decreased to $82\%$ for $\hbar\gamma^{(m)}=0$~$\mu$eV at $\hbar\gamma_d^{(m)}=1$~$\mu$eV and to $\sim70\%$ for state-of-art InGaAs QD excitonic lifetimes. For very large intercavity distances, where the splitting is of the order of the normal mode linewidth, the entanglement is more sensible and it is decreased to $\sim50\%$ at $\hbar\gamma_d^{(m)}=1$~$\mu$eV for realistic InGaAs QDs. Since we are interested in the strong cavity-cavity coupling regime, i.e., well-defined normal mode splitting, and low-loss QD excitonic states, we will safely consider $\gamma_d^{(m)}=0$ in the calculations below. Equivalent results were obtained for entanglement as a function of the pure dephasing rates in the $0^{\circ}$, $60^{\circ}$, and $90^{\circ}$ dimers. As a final remark on the investigation of the main dissipation sources in the system, by coherently pumping the QDs it is possible to produce a residual incoherent pumping of the resonant normal mode. However, since we are in the weak pumping regime, we verified (not shown here) that this unwanted effect, modeled by an additional term  ${\cal L}(P)=P\left[\hat{a}^{\dag}\tilde{\rho}\hat{a}-\hat{a}\hat{a}^{\dag}\tilde{\rho}/2-\tilde{\rho} \hat{a}\hat{a}^{\dag}/2\right]$ in Eq.~(\ref{master}) with a rate up to $P=2\Omega$, modifies the amount of the entanglement by only a few percent and can be safely neglected.

\begin{figure}[t!]
  \begin{center}
    \includegraphics[width=0.45\textwidth]{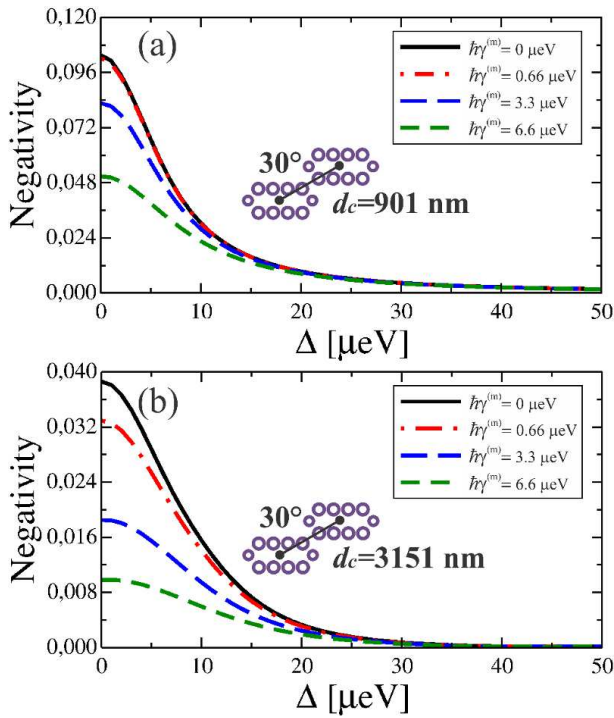}
  \end{center}
  \caption{
(Color online) Steady-state negativity for the $30^{\circ}$ dimer at $d_c=2\sqrt{3}a=901$~nm, panel (a), and $d_c=7\sqrt{3}a=3151$~nm, panel (b), as a function of the detuning between the excitonic transition frequencies of the QDs, for different QD loss rates. The QD 1 is in resonance with the lower frequency normal mode, i.e., $\omega^{(1)}=\omega_1$, and $\omega^{(2)}=\omega^{(1)}+\Delta$ with a pumping rate $\hbar\Omega_0=1$~$\mu$eV.}
\label{fig:neg-vs-det}
\end{figure}

Since QDs are very likely to be detuned due to their inhomogeneous size distribution, we finally studied the effects of the differences of QD excitonic transition frequencies on the steady-state entanglement. The results for the $30^{\circ}$ dimer, at the minimum and maximum interdot distances, are shown in Figs.~\ref{fig:neg-vs-det}(a) and \ref{fig:neg-vs-det}(b), respectively. The same QD loss rates of Fig.~\ref{fig:neg-vs-d} were considered here, but neglecting pure dephasing. The entanglement is a very sensitive function of QDs detuning. In fact, the negativity drops from 20$\%$ (at $\Delta=0$) to 5$\%$ of the maximal value for detuning $\Delta=10$~$\mu$eV, see Fig.~\ref{fig:neg-vs-det}(a), and from 8$\%$ ($\Delta=0$) to 3$\%$ of the maximum negativity in Fig.~\ref{fig:neg-vs-det}(a). The presence of the second normal mode at large intercavity distances explains the smoother decreasing in the curves of Fig.~\ref{fig:neg-vs-det}(b) as compared to the corresponding curves in Fig.~\ref{fig:neg-vs-det}(a). Radiative coupling between the QDs through a photonic normal mode of the PCD is possible as long as the non-resonant condition determines a detuning between QDs that is smaller than the mode linewidth; nevertheless, Fig.~\ref{fig:neg-vs-det} evidences that the condition for entanglement between radiatively-coupled QDs is more stringent. In Fig.~\ref{fig:neg-vs-det}(a), the linewidth of the corresponding photonic normal mode is $67$~$\mu$eV, but the negativity is close to zero for detuning values larger than $40$~$\mu$eV, where an effective radiative coupling is still present between the QDs. Hence, the entanglement is more conditioned by the linewidth of the excitonic transitions than the linewidth of the coupled photonic normal mode, meaning that the entanglement is sizable only when the QDs detuning is smaller than their linewidth. We have obtained equivalent results for the $0^{\circ}$, $60^{\circ}$, and $90^{\circ}$ PCD, respectively (results not shown).

\section{Entanglement Dynamics} \label{DyEnt}

\begin{figure}[pt!]
  \begin{center}
    \includegraphics[width=0.45\textwidth]{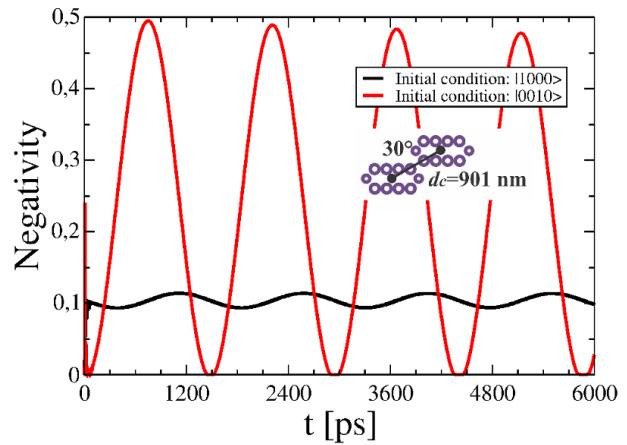} 
  \end{center}
  \caption{
(Color online) Negativity dynamics in the $30^{\circ}$ dimer at $d_c=2\sqrt{3}a=901$~nm and $\gamma^{(m)}=0$, considering the initial conditions $|1000\rangle$, in black, and $|0010\rangle$, in red.}
\label{fig:neg-vs-t-dc1}
\end{figure}

In Sec.~\ref{SSEnt}, we showed that it is possible to obtain 20$\%$ of the maximum entanglement between two radiatively-coupled QDs in the steady-state regime after resonant and continuous-wave driving of the fundamental excitonic transitions, for interdot separations that can be sizably larger than the characteristic operational wavelength of the system. However, practical applications for quantum information technologies require strongly entangled qubits.
In this respect, our scheme for steady-state entanglement might still be optimized. One possibility would be to  consider an asymmetric pump/detection configuration as, e.g., in Ref.~\onlinecite{Elena}, and make a global optimization search in the parameters' space, which goes beyond the scope of the present work.
On the other hand, an immediate application of the model employed here could allow for larger values of the negativity to  be achieved in the transient dynamics, as also pointed out in the literature \cite{hughes4,gonzalez2011prl}. In this section, we focus on the $30^{\circ}$ PCD, which is the most convenient configuration for entanglement applications, and we consider the two QDs in resonance with the lower frequency normal mode. We assume $\hbar\Omega_0=1$~$\mu$eV at the optimal phase difference between the QD coherent drivings. The basis $|\alpha_1 \alpha_2 m_1 m_2\rangle$ for solving the dynamics of the master equation, Eq.~(\ref{master}), is safely truncated at $m_i=1$ (previously checked for convergence), as in Sec.~\ref{SSEnt}. Figure~\ref{fig:neg-vs-t-dc1} shows the negativity dynamics up to 6~ns at $d_c=2\sqrt{3}a=901$~nm and $\gamma^{(m)}=0$, for two different initial conditions: a single excitation in QD 1, i.e., initial state $|1000\rangle$, and a single photon in the lower frequency normal mode, i.e., initial state $|0010\rangle$, respectively. The negativity oscillates with a frequency determined by the pumping rate, $\sim\Omega_0/2$, and the amplitude of the oscillations approximates the maximum negativity value of 0.5, when the initial excitation is in the photonic mode. The latter is the most favorable situation, since the two QDs are equally populated in time by the field, giving rise to an optimal condition for maintaining the entanglement through the resonant-QD coherent pumping. When we consider an excited QD at $t=0$, the two QDs are not equally populated in time, which yields an unfavorable condition for their mutual entanglement. As it is physically expected, the amplitude of the oscillations decreases with increasing time, due to the normal mode dissipation, tending asymptotically to the steady-state negativity. 

\begin{figure}[pt!]
  \begin{center}
    \includegraphics[width=0.45\textwidth]{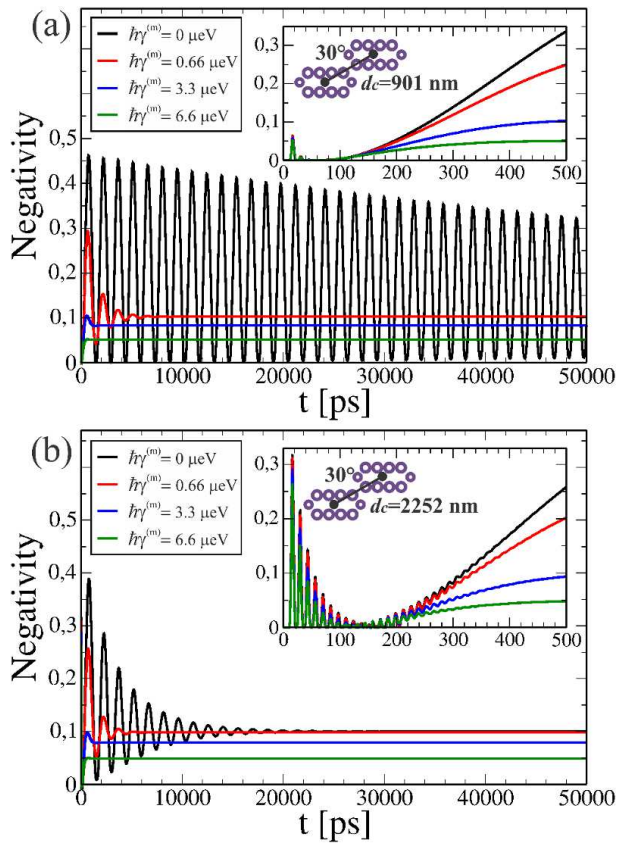}
  \end{center}
  \caption{
(Color online) Numerical experiment for time-dependent negativity in the transient dynamics using the $30^{\circ}$ dimer at $d_c=2\sqrt{3}a=901$~nm, panel (a), and $d_c=5\sqrt{3}a=2252$~nm, panel (b). The initial condition is $|1000\rangle$ in both cases with QD 1 and photonic mode $\omega_1$ in resonance, and QD 2 far from resonance. We wait for a time $\tau=9$~ps and $\tau=9.3$~ps, at $d_c=901$~nm and $d_c=2252$~nm, respectively, with the aim of maximally populating the resonant photonic mode, and QD 2 is brought into resonance for $t>\tau$. The QD pumping rate is $\hbar\Omega_0=1$~$\mu$eV at resonance with the excitonic dark state for all times. The insets show the early dynamics of the system.}
\label{fig:neg-vs-t}
\end{figure}

The results reported in Fig.~\ref{fig:neg-vs-t-dc1} show that the optimal initial condition is given by a single excitation in the photonic mode, while the QDs are in their ground state at $t=0$. Nevertheless, this is particularly challenging due to the delocalized nature of the normal mode: in order to achieve such an initial condition, it would be necessary to prepare a collective state of both cavities at the same time. Here, we propose a different and less challenging operational approach. We consider QD 2 initially out of resonance, and QD 1 in resonance with the normal mode at frequency $\omega_1$, and we assume an initial condition with a single exciton in QD 1 at $t=0$, i.e., $|1000\rangle$. Then, we wait for a time $\tau$ at which the excitation has been almost completely transferred to normal mode 1 due to Rabi coupling. At this time, QD 2 is brought into resonance with QD 1, which can be accomplished by using the quantum confined Stark effect \cite{villasboas,faraon}, for example. All these steps are performed by pumping the QDs at the  frequency of the dark state of the system. The results of this numerical experiment are shown in Fig.~\ref{fig:neg-vs-t}(a) for the $30^{\circ}$ PCD at $d_c=2\sqrt{3}a=901$~nm separation, and in Fig.~\ref{fig:neg-vs-t}(b) for the same PCD configuration at $d_c=5\sqrt{3}a=2252$~nm center-to-center distance. Here we have assumed the same QD loss rates as in the calculations of the previous Section, and the intercavity distances considered here delimit the flat region in Fig.~\ref{fig:neg-vs-d}(a). From Fig.~\ref{fig:neg-vs-t}(a), it is evident that our approach is totally equivalent to consider an initial excitation in the photonic mode, and the amount of entanglement is also very close to the maximum value obtained in the transient dynamics for $\gamma^{(m)}=0$. As in Fig.~\ref{fig:neg-vs-t-dc1}, the amplitude of the oscillations decreases with increasing time due to dissipation in the resonant normal mode of the PCD. When QD losses are taken into account, maximum negativity values around $\sim0.2$, i.e., $40\%$ of the maximal value, are obtained for state-of-art QDs. However, the presence of this dissipation channel produces a faster decreasing amplitude as compared to the corresponding result for $\gamma^{(m)}=0$. As a consequence, the steady-state value is achieved more rapidly. For interdot distance $d_c=2252$ nm, i.e., the results shown in Fig.~\ref{fig:neg-vs-t}(b), the presence of the second normal mode, with the same symmetry of the excitonic dark state, starts to play a role in the transient dynamics, providing an additional loss channel for the entangled QDs; even for $\gamma^{(m)}=0$, the steady-state regime is rapidly achieved. Nevertheless, maximum negativity values of about $\sim0.2$, or $40\%$ of the maximal value, are obtained for state-of-art QDs. In the early dynamics, as shown in insets of Fig.~\ref{fig:neg-vs-t}, the fast oscillation frequency is determined by the QD-PCD mode coupling rates, $g_m^{(n)}$, and the negativity amplitude is affected by the loss rates: Since the normal mode losses, $\gamma_m$, are smaller for the $d_c=2252$~nm than for the $d_c=901$~nm intercavity distance, the negativity amplitude is larger for $d_c=2252$~nm than for $d_c=901$~nm. However, the slow transient dynamics (i.e., after 200~ps) determines large negativity time intervals that are much larger than the photonic mode and QD lifetimes, which could be relevant for practical applications of transient QDs entanglement.

\section{Conclusions} \label{Concl}

We have studied the conditions for achieving steady-state entanglement between radiatively-coupled quantum dots by exploiting their mutual long-distance interaction through the normal modes of a photonic crystal dimer. The amount of entanglement is quantified through the Peres-Horodecki negativity criterion of the reduced density matrix within the two QD subspace, which is computed through the quantum-dissipative master equation in the Markov approximation. The photonic crystal slab structures were solved within the guided mode expansion approach, and their solutions were used as input parameters for the master equation formalism. Material parameters relevant to InGaAs and GaAs nanostructures were considered throughout the work, but the results can be generalized to an arbitrary material platform.

In the steady-state regime and for resonant pumping condition, we have found that the largest entanglement is obtained at the excitonic dark state of the system, i.e., for a coherent driving of the quantum dots with a phase difference of (a) $(2n+1)\pi$ when coupled to a symmetric normal mode of the photonic crystal dimer, and (b) $2n\pi$ when they are coupled to an antisymmetric one, respectively, $n$ being an integer. The largest negativity value achieved in this regime is predicted to be on the order of $\sim0.1$, i.e., $20\%$ of the maximum value, and it remains of the same order of magnitude as long as the normal mode splitting is well-defined, i.e., larger than the photonic mode linewidths. These results are shown to be robust against the main sources of QD decoherence, such as spontaneous emission and pure dephasing.
Furthermore, when the splitting is of the order of the photonic mode linewidths, the negativity is roughly proportional to the normal mode splitting. On the other hand, when a QDs inhomogeneous distribution is considered, the entanglement is shown to remain sizable only for detunings that are smaller than their linewidths. As a consequence, the QD radiative coupling is a necessary but not a sufficient condition to obtain entanglement between the two qubits. 
In terms of the photonic crystal dimer, our results show that the $30^{\circ}$ dimer is the most convenient configuration to show long-range entanglement, due to its very-well-defined normal mode splitting even at intercavity distances that are larger than the characteristic operational wavelength of the system. 

When addressing the transient dynamics of the system, it has been shown that the degree of entanglement can be sizably larger than the steady-state value. In such a case, we found that an optimal condition for initializing the system is obtained when considering an initial excitation in the resonant normal mode, where long-time negativity oscillations with a frequency $\sim\Omega_0/2$ are seen with a period much larger than both photon and exciton lifetimes, respectively. Based on these results, we have proposed and demonstrated an effective protocol for generating the same long-time entanglement oscillations in practical devices, by initializing the system with a single excitation in one of the quantum dots (which represents an operationally less challenging task). Negativity values of the order of $\sim 0.2$, i.e., $40\%$ of the maximum value, were obtained for state-of-art InGaAs quantum dots in our proposed device. As a final remark, we believe that the present system will be useful for quantum information applications on photonic crystal platforms, where the entanglement between distant qubits is a key functionality to be developed.

\section{Acknowledgements}
We acknowledge financial support from Brazilian funding agencies FAPEMIG, CAPES, and CNPq. MFS thanks CNPq (305384/2015-5). This work was partially financed by PVE-Ci\^{e}ncia Sem Fronteiras/CNPq Project No. 407167/2013-7.

\appendix

\section{Negativity of Bell states}\label{appx}
We hereby discuss the upper bound for the negativity value of two maximally entangled qubits.
The Bell states can be written in the two-qubit basis as follows:
\begin{align}
 |\phi^{\pm}\rangle&=\frac{1}{\sqrt{2}}(|00\rangle\pm|11\rangle),\\
 |\psi^{\pm}\rangle&=\frac{1}{\sqrt{2}}(|01\rangle\pm|10\rangle),
\end{align}
where the corresponding density operators are given by
\begin{align}
\hat{\rho}_{\phi^{\pm}}=|\phi^{\pm}\rangle\langle\phi^{\pm}|, & & \hat{\rho}_{\psi^{\pm}}=|\psi^{\pm}\rangle\langle\phi^{\pm}| \label{oprho}
\end{align}
Considering the ordering of the basis $\left\{|00\rangle,|01\rangle,|10\rangle,\right.$  $\left.|11\rangle\right\}$, the matrix representations of the density operators in Eq.~(\ref{oprho}) read
\begin{align}
\rho_{\phi^{\pm}}=\frac{1}{2}
\begin{pmatrix}
 1 & 0 & 0 & \pm1 \cr
 0 & 0 & 0 & 0 \cr
 0 & 0 & 0 & 0 \cr
 \pm1 & 0 & 0 & 1
\end{pmatrix},
& &
\rho_{\psi^{\pm}}=\frac{1}{2}
\begin{pmatrix}
 0 & 0 & 0 & 0 \cr
 0 & 1 & \pm1 & 0 \cr
 0 & \pm1 & 1 & 0 \cr
 0 & 0 & 0 & 0
\end{pmatrix}.\label{mrho}
\end{align}
The matrix elements of $\rho^{T1}$, namely, the partial transpose of $\rho$ with respect to qubit 1, i.e., the first entry of $|\alpha_1\alpha_2\rangle$, are obtained from the matrix elements of $\rho$ following the rule $\langle\alpha_1\alpha_2|\rho^{T_1}|\alpha_1'\alpha_2'\rangle=\langle\alpha_1'\alpha_2|\rho|\alpha_1\alpha_2'\rangle$. The matrix representations of $\rho_{\phi^{\pm}}^{T1}$ and $\rho_{\psi^{\pm}}^{T1}$ are then 
\begin{align}
\rho_{\phi^{\pm}}^{T1}=\frac{1}{2}
\begin{pmatrix}
 1 & 0 & 0 & 0 \cr
 0 & 0 & \pm1 & 0 \cr
 0 & \pm1 & 0 & 0 \cr
 0 & 0 & 0 & 1
\end{pmatrix},
& &
\rho_{\psi^{\pm}}^{T1}=\frac{1}{2}
\begin{pmatrix}
 0 & 0 & 0 & \pm1 \cr
 0 & 1 & 0 & 0 \cr
 0 & 0 & 1 & 0 \cr
 \pm1 & 0 & 0 & 0
\end{pmatrix}.\label{mrhoT1}
\end{align}
Finally, it is easy to show that the characteristic equation to find the eigenvalues $\lambda$ of the four matrices in Eq.~(\ref{mrhoT1}) is
\begin{equation}
\left(0.5-\lambda\right)^3\left(0.5+\lambda\right)=0, \label{chareqn}
\end{equation}
and their solutions are $\{0.5,0.5,0.5,-0.5\}$. The absolute value of the sum of the negative eigenvalues in Eq.~(\ref{chareqn}), i.e, the negativity, is therefore $0.5$. Since Bell states are maximally entangled and the negativity is an entanglement monotone for composite Hilbert spaces of dimension $2\otimes2$, a negativity value of $0.5$ determines an upper bound for the amount of entanglement in a two-qubit system.


\end{document}